\documentclass[aps, prl, twocolumn, superscriptaddress]{revtex4}

\usepackage[dvips]{graphicx}
\usepackage{color}
\usepackage{amssymb}
\renewcommand{\vec}[1]{\boldsymbol{#1}}
\usepackage{dcolumn}
\usepackage{bm}
\usepackage{times}

\usepackage{chemformula} 
\begin{document}
\newcommand{\veck}[1]{\boldsymbol{#1}}

\title{Realization of the orbital-selective Mott state at the molecular level in \ch{Ba3LaRu2O9}}

\author{Q. Chen}
\affiliation{Department of Physics and Astronomy, University of Tennessee, Knoxville, Tennessee 37996, USA}

\author{A. Verrier}
\affiliation{Institut Quantique and Departement de Physique, Universite de Sherbrooke, Sherbrooke, Quebec J1K 2R1, Canada}

\author{D. Ziat}
\affiliation{Institut Quantique and Departement de Physique, Universite de Sherbrooke, Sherbrooke, Quebec J1K 2R1, Canada}

\author{A.J. Clune}
\affiliation{Department of Chemistry, University of Tennessee, Knoxville, Tennessee 37996, USA}

\author{R. Rouane}
\affiliation{Institut Quantique and Departement de Physique, Universite de Sherbrooke, Sherbrooke, Quebec J1K 2R1, Canada}

\author{X. Bazier-Matte }
\affiliation{Institut Quantique and Departement de Physique, Universite de Sherbrooke, Sherbrooke, Quebec J1K 2R1, Canada}

\author{G. Wang}
\affiliation{Key Laboratory of Artificial Structures and Quantum Control, School of Physics and Astronomy, Shanghai Jiao Tong University, Shanghai 200240, China}

\author{S. Calder}
\affiliation{Neutron Scattering Division, Oak Ridge National Laboratory, Oak Ridge, Tennessee 37831, USA}

\author{K.M. Taddei}
\affiliation{Neutron Scattering Division, Oak Ridge National Laboratory, Oak Ridge, Tennessee 37831, USA}

\author{C. R. dela Cruz}
\affiliation{Neutron Scattering Division, Oak Ridge National Laboratory, Oak Ridge, Tennessee 37831, USA}

\author{A.I. Kolesnikov}
\affiliation{Neutron Scattering Division, Oak Ridge National Laboratory, Oak Ridge, Tennessee 37831, USA}

\author{J. Ma}
\affiliation{Key Laboratory of Artificial Structures and Quantum Control, School of Physics and Astronomy, Shanghai Jiao Tong University, Shanghai 200240, China}

\author{J.-G. Cheng}
\affiliation{Beijing National Laboratory for Condensed Matter Physics and Institute of Physics, Chinese Academy of Sciences, Beijing 100190, China}
\affiliation{Songshan Lake Materials Laboratory, Dongguan, Guangdong 523808, China}

\author{Z. Liu}
\affiliation{Department of Physics, University of Illinois at Chicago, Illinois 60607, USA}

\author{J. A. Quilliam}
\affiliation{Institut Quantique and Departement de Physique, Universite de Sherbrooke, Sherbrooke, Quebec J1K 2R1, Canada}

\author{J.L. Musfeldt}
\affiliation{Department of Physics and Astronomy, University of Tennessee, Knoxville, Tennessee 37996, USA}
\affiliation{Department of Chemistry, University of Tennessee, Knoxville, Tennessee 37996, USA}

\author{H.D. Zhou}
\affiliation{Department of Physics and Astronomy, University of Tennessee, Knoxville, Tennessee 37996, USA}

\author{A.A. Aczel}
\email[Email to: ]{aczelaa@ornl.gov}
\affiliation{Department of Physics and Astronomy, University of Tennessee, Knoxville, Tennessee 37996, USA}
\affiliation{Neutron Scattering Division, Oak Ridge National Laboratory, Oak Ridge, Tennessee 37831, USA}

\date{\today}

\begin{abstract}	
Molecular magnets based on heavy transition metals have recently attracted significant interest in the quest for novel magnetic properties. For systems with an odd number of valence electrons per molecule, high or low molecular spin states are typically expected in the double exchange or quasi-molecular orbital limits respectively. In this work, we use bulk characterization, muon spin relaxation, neutron diffraction, and inelastic neutron scattering to identify a rare intermediate spin-3/2 per dimer state in the 6H-perovskite Ba$_3$LaRu$_2$O$_9$ that cannot be understood in a double exchange or quasi-molecular orbital picture and instead arises from orbital-selective Mott insulating behavior at the molecular level. Our measurements are also indicative of collinear stripe magnetic order below $T_N$~$=$~26(1)~K for these molecular spin-3/2 degrees-of-freedom, which is consistent with expectations for an ideal triangular lattice with significant next nearest neighbor in-plane exchange. Finally, we present neutron diffraction and Raman scattering data under applied pressure that reveal low-lying structural and spin state transitions at modest pressures P~$\le$~1\,GPa, which highlights the delicate balance between competing energy scales in this system.
\end{abstract}

\maketitle

\section{Introduction}

The interplay between charge, spin, lattice and orbital degrees of freedom leads to a tremendous variety of exotic phenomena in strongly-correlated electron systems. One particularly famous example is the metal-Mott insulator transition, where a Coulomb repulsion $U$ that is strong relative to orbital hopping $t$ leads to a significant modification of the band structure and promotes complete electron localization and hence local moment physics. More recently, the intriguing concept of an orbital-selective Mott phase was proposed in order to explain the coexistence of itinerant and localized electron character in the ruthenate system Sr$_\text{2-x}$Ca$_\text{x}$RuO$_4$ \cite{02_anisimov}. This state can be achieved in multi-orbital systems with disparate orbital hoppings, leading to a situation where some valence electrons are localized while others are itinerant. Although the realization of the orbital-selective Mott phase is still under debate for Sr$_\text{2-x}$Ca$_\text{x}$RuO$_4$ \cite{04_wang, 05_balicas, 07_liebsch, 09_neupane, 09_demedici}, this state has been the subject of several theoretical investigations \cite{04_koga, 05_demedici, 05_ferrero, 05_liebsch, 05_biermann, 14_rincon_1, 14_rincon_2} and has been discussed in the context of iron-based superconductivity \cite{12_yu, 13_yu}, high-$T_c$ cuprates \cite{02_venturini}, the metal-insulator transition in V$_2$O$_3$ \cite{06_laad}, and the magnetic properties of double perovskites \cite{18_chen} and BaFe$_2$Se$_3$ \cite{15_mourigal, 18_herbrych}. 

Orbital-selective Mott physics may also play an important role in heavy transition metal (i.e. 4$d$ and 5$d$) molecular magnets \cite{16_streltsov, 17_streltsov}. The large spatial extent of the $d$ orbitals and the reduction of Hund's coupling $J_H$ generate competing energy scales that can invalidate the local moment / double exchange picture generally expected in their 3$d$ transition metal counterparts and therefore have significant consequences on their electronic ground states and magnetic properties. For a system with two types of orbitals (c and d), three regimes are possible depending on the relative strengths of $t_c$, $t_d$, $J_H$, and $U$: (i) the quasi-molecular orbital limit ($t_c$, $t_d$~$>>$~$J_H$, $U$), (ii) the local moment / double exchange limit ($t_c$, $t_d$~$<<$~$J_H$, $U$), and (iii) the orbital-selective limit ($t_c$~$>>$~$J_H$, $U$, $t_d$~$\rightarrow$~0) \cite{14_streltsov}. An accurate determination of the electronic ground state for these molecular magnets is the first step towards developing a detailed understanding of their collective magnetic properties, which may be quite interesting if the molecules are strongly-interacting. 

Heavy transition metal molecular magnets with an odd number of electrons per dimer are an excellent playground for identifying systems and establishing trends where the double exchange picture breaks down at the molecular level. The competition between Hund’s coupling and orbital hopping is particularly striking in this case, as the electronic ground state evolves from high to low spin with increasing orbital hopping \cite{16_streltsov}. In principle, for particular electron configurations an intermediate spin ground state can also be realized in the orbital-selective regime but experimental examples are lacking. This class of materials may therefore enable detailed investigations of low-lying spin state transitions, which are not common in magnetic materials based on single ion building blocks. Furthermore, in the quasi-molecular orbital limit they can also generate new $S$~$=$~1/2 quantum magnets with ideal frustrated lattice geometries that are less susceptible to Jahn-Teller distortions \cite{17_ziat} and in some cases are under active consideration as quantum spin liquid candidates \cite{12_sheckelton, 14_mourigal, 18_akbari}. 

\begin{figure*}[ht]
	\centering
	\scalebox{0.5}{\includegraphics{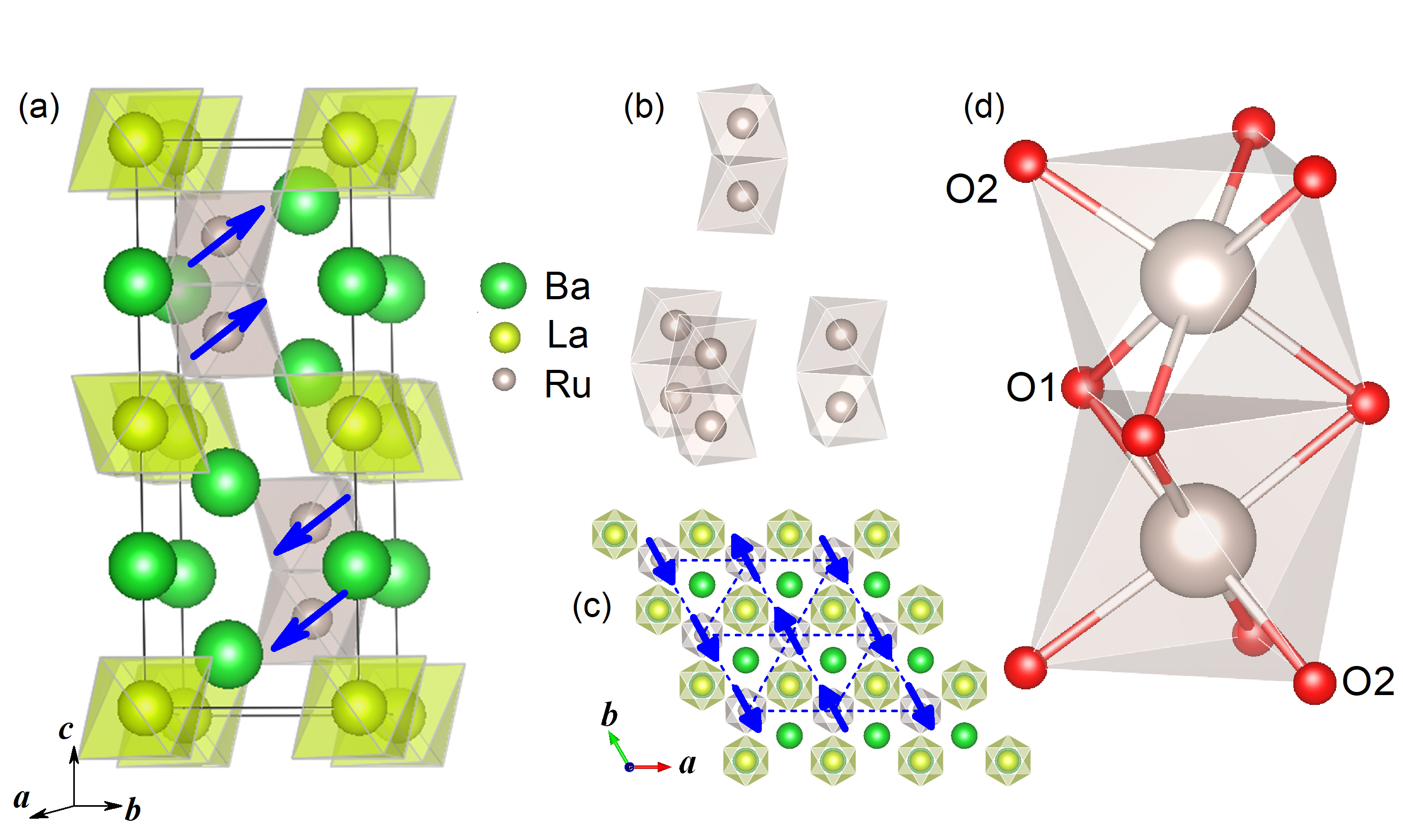}}
	\caption{(color online) (a) A crystallographic unit cell of \ch{Ba3LaRu2O9}, with the oxygen atoms excluded for clarity. The blue arrows represent the magnetic structure determined by neutron powder diffraction. (b) A schematic of the Ru dimers in \ch{Ba3LaRu2O9}. (c) A view along the c-axis of the triangular lattice formed by the Ru dimers in the cystallographic $ab$-plane. (d) A close-up view of one Ru dimer in \ch{Ba3LaRu2O9}, which illustrates the face-sharing nature of the octahedra and the positions of the two crystallographically-inequivalent oxygen ions.}
	\label{Fig1}
\end{figure*}

The 6H-perovskites, with the general chemical formula Ba$_3$\textit{MR}$_2$O$_9$, consist of transition metal dimers decorating a triangular lattice as illustrated in Fig.~\ref{Fig1} and they have already been shown to host a variety of interesting electronic ground states at the molecular level. For example, Ba$_3$NaRu$_2$O$_9$ exhibits interdimer charge order below 210~K \cite{12_kimber}, Ba$_3$(Y,In,Lu)Ru$_2$O$_9$ are quantum magnets with the $S_\text{tot}$~$=$~1/2 degree of freedom delocalized over the Ru dimers \cite{17_ziat}, and Ba$_3$CeRu$_2$O$_9$ has a non-magnetic ground state that arises from quasi-molecular orbital formation combined with a large zero field splitting \cite{19_chen}. The rich molecular behavior in this family likely arises from the ability of this structure to accommodate heavy transition metal dimers based on face-sharing octahedra, which feature metal-metal distances shorter than the nearest neighbor distance in the corresponding elemental metal \cite{19_chen} in some cases. In fact, quasi-molecular orbital formation has been argued to give rise to the quantum magnetism in Ba$_3$(Y,In,Lu)Ru$_2$O$_9$ \cite{17_ziat} rather than the large spin state of $S_\text{tot}$~$=$~5/2 per dimer that would be realized by a double exchange mechanism \cite{10_bechlars}. 

Interestingly, despite the same valence electron count of seven per Ru dimer, the magnetic properties of the isostructural system Ba$_3$LaRu$_2$O$_9$ have been shown to be drastically different. The effective and ordered moments per dimer, extracted via magnetic susceptibility \cite{02_doi} and neutron diffraction \cite{13_senn} measurements respectively, are much larger and cannot be explained by an $S_\text{tot}$~$=$~1/2 electronic ground state for the dimers. No satisfactory explanation for this different behavior has been proposed to-date and the true electronic ground state of the Ru dimers in Ba$_3$LaRu$_2$O$_9$ has remained an open question. In this work, we combine magnetometry, heat capacity, muon spin relaxation, neutron diffraction, and inelastic neutron scattering to identify an $S_\text{tot}$~$=$~3/2 electronic ground state for the dimers that we argue arises from an orbital-selective mechanism at the molecular level. We also establish an ordering temperature of $T_N$~$=$~26(1)~K for the stripe spin configuration that is expected for a triangular lattice with a significant next-nearest neighbor in-plane exchange interaction. Finally, we use neutron powder diffraction and Raman scattering under applied pressure to show that moderate pressures of $P$~$\le$~1 GPa generate both structural and spin state transitions in Ba$_3$LaRu$_2$O$_9$. 


\section{Experimental methods}

Polycrystalline samples of \ch{Ba3LaRu2O9} were synthesized by a solid-state reaction using a stoichiometric amount of the starting materials \ch{BaCO3}, Ru, and \ch{La2O3} (fine powder predried at 950\textdegree C overnight) with purities of 99.9\% or higher. The starting materials were mixed in agate mortars, pressed into pellets, annealed in air at 900\textdegree C for 12 hours, and then annealed at 1200\textdegree C for 20 hours with intermediate grinding and pelletizing. 

The room temperature X-ray powder diffraction (XRD) patterns were collected using a HUBER Image Plate Guinier Camera 670 with Ge monochromatized Cu $K_{\alpha1}$ radiation ($\lambda \approx 1.54$\,\text{\AA}) to check the quality of the powder samples. No obvious impurity peaks were observed.

The dc magnetic susceptibility and magnetization measurements were performed in the temperature range of 2-320\,K using a Quantum Design superconducting interference device (SQUID) magnetometer. The high-temperature magnetic susceptibility was measured with a Quantum Design Magnetic Property Measurement System (MPMS) in the temperature range of 300-800\,K. The specific heat measurements were performed using the relaxation method with a commercial Physical Property Measurement System (PPMS) from Quantum Design.

Neutron powder diffraction (NPD) was performed on $\sim$~6.5 g of polycrystalline \ch{Ba3LaRu2O9} using the HB-2A powder diffractometer of the High Flux Isotope Reactor (HFIR) at Oak Ridge National Laboratory (ORNL) \cite{18_calder}. The sample was loaded in a cylindrical vanadium can, and the data were collected at different temperatures ranging from 1.5~K to 300~K with neutron wavelengths of 1.54~\AA~and 2.41~\AA~and a collimation of open-21$'$-12$'$. The ambient pressure HB-2A data was refined using the FullProf software suite \cite{93_rodriguez} and the magnetic structure symmetry analysis was performed using SARAh \cite{00_wills}. Further NPD studies were also carried out on HB-2A with $\sim$~4.5 g of polycrystalline \ch{Ba3LaRu2O9} and a Fluorinert pressure medium first loaded in a teflon tube and then placed in a Cu-Be clamp cell capable of applying hydrostatic pressures up to 1 GPa. Elastic neutron scattering measurements, complementary to the ambient pressure NPD experiment described above, were performed on the 14.6~meV fixed-incident-energy triple-axis spectrometer HB-1A of the HFIR at ORNL using the same polycrystalline sample of \ch{Ba3LaRu2O9} measured in ambient pressure at HB-2A over a temperature range 1.5~K to 40~K. For this experiment, the sample was loaded in a cylindrical Al can to minimize incoherent nuclear scattering that could prevent the detection of weak magnetic Bragg peaks not observed in the initial HB-2A experiment. The overall background was minimized by using a double-bounce monochromator system, mounting two-highly oriented pyrolytic graphite (PG) filters in the incident beam to remove higher-order wavelength contamination, and placing an analyzer of PG crystals before the single He-3 detector for energy discrimination. A collimation of 40$'$-40$'$-40$'$-80$'$ resulted in an energy resolution at the elastic line just over 1 meV (FWHM). 

Inelastic neutron-scattering (INS) measurements were performed on the direct-geometry time-of-flight chopper spectrometer SEQUOIA \cite{10_granroth} of the Spallation Neutron Source (SNS) at ORNL, using the same \ch{Ba3LaRu2O9} polycrystalline measured in the ambient pressure HB-2A experiment. The sample was loaded in a cylindrical Al can and spectra were collected with incident energies $E_i$~$=$~25 and 100~meV at temperatures of 4~K (both incident energies), 30 K (25 meV only), and 300 K (100 meV only). An empty aluminum can was measured in identical experimental conditions for a similar counting time. The resulting background spectra were subtracted from the corresponding sample spectra after normalization with a vanadium standard to account for variations of the detector response and the solid angle coverage.

Muon spin relaxation ($\mu$SR) measurements were performed on the M20 surface muon beamline at TRIUMF. A low-background ``veto'' set-up was employed with the sample mounted inside a mylar packet placed in the path of the muon beam within a helium flow cryostat. Measurements were performed in zero field and longitudinal field geometries over a temperature range of 1.5 to 60~K. A good review of the $\mu$SR technique can be found in Ref.~\cite{11_yaouanc}.

Raman scattering was performed under compression between 0 to 10.36 GPa using diamond anvil cell techniques. Polycrystalline material was loaded into a symmetric diamond anvil cell along with an annealed ruby ball, and KBr was used as the pressure medium for the measurement. This assured a quasi-hydrostatic environment for the sample. Fluorescence from the ruby ball was used to determine pressure \cite{86_mao}. These experiments  were carried out using the COMPRES beamline facility at the National Synchrotron Light Source II at Brookhaven National Laboratory. We employed $\lambda_{excit}$ = 532 nm; $\approx$1 mW power; 30 sec integration, averaged three times. All data were collected at room temperature.


\begin{figure*}[ht]
	\centering 
	\scalebox{0.65}{\includegraphics{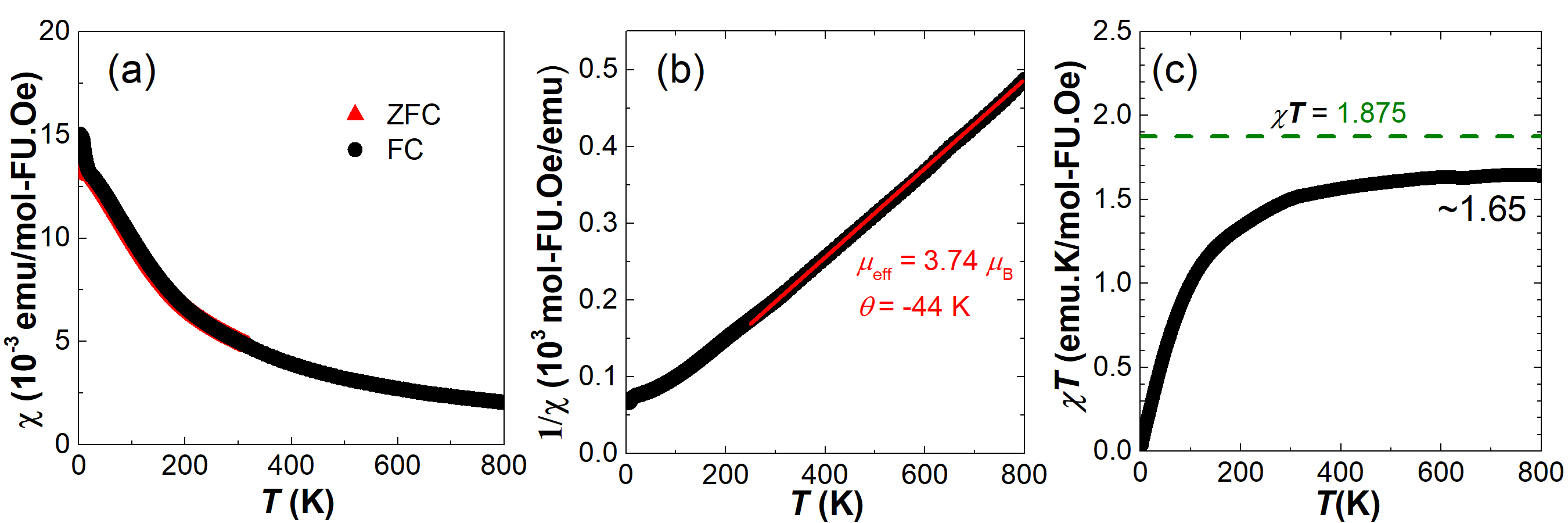}}
	\caption{(color online) (a) Magnetic susceptibility vs temperature for \ch{Ba3LaRu2O9} measured in an applied magnetic field of 1 kOe under both zero-field-cooled and field-cooled conditions between 2 and 800 K. (b) Inverse susceptibility vs temperature with the Curie-Weiss fit indicated by the solid red line. (c) The temperature dependence of $\chi T$ reaches a saturation value of $\sim$1.65, which is close to the expected value of $\chi T = 1.875$ indicated by the dashed line for $S_{tot}$~$=$~3/2 and $g$~$=$~2.}
	\label{Fig2}
\end{figure*}

\section{Results and Discussion}

\subsection{(I) Electronic ground state of the Ru dimers} 
	
Figure~\ref{Fig2}(a) shows the dc magnetic susceptibility $\chi$ (plotted as $M/H$) vs temperature for \ch{Ba3LaRu2O9} measured from 2 to 800\,K in an applied magnetic field of 1\,kOe under both zero-field cooled (ZFC) and field-cooled (FC) conditions. The low-temperature data below $\sim$~25 K is indicative of magnetic order and will be discussed more in the next section. The Curie-Weiss fit of the high-temperature inverse susceptibility data (above 250\,K) is shown in Fig.~\ref{Fig2}(b). This fit results in a Curie-Weiss temperature $\theta_\text{CW} = -44$\,K and an effective moment $\mu_\text{eff} = 3.74\,\mu_B$/FU (\ch{Ba3LaRu2O9} formula unit), which is close to the expected value of 3.87$\,\mu_B$/dimer for a molecular spin-3/2 state. The $\chi T$ versus $T$ plot shown in Fig.~\ref{Fig2}(c) reaches a saturation value of $\sim$\,1.65 that is slightly below $\chi T=1.875$ for spin-3/2, which could be due to the thermal population of spin-1/2 excited states in the high-temperature region (up to 800\,K). This suggests that the Ru-dimers in \ch{Ba3LaRu2O9} adopt an unusual spin-3/2 ground state, in sharp contrast to the isostructural analogs Ba$_3$(Y,In,Lu)Ru$_2$O$_9$ that are known to host spin-1/2 dimer ground states \cite{17_ziat}.    

\begin{figure*}[ht]
\centering 
\scalebox{0.65}{\includegraphics{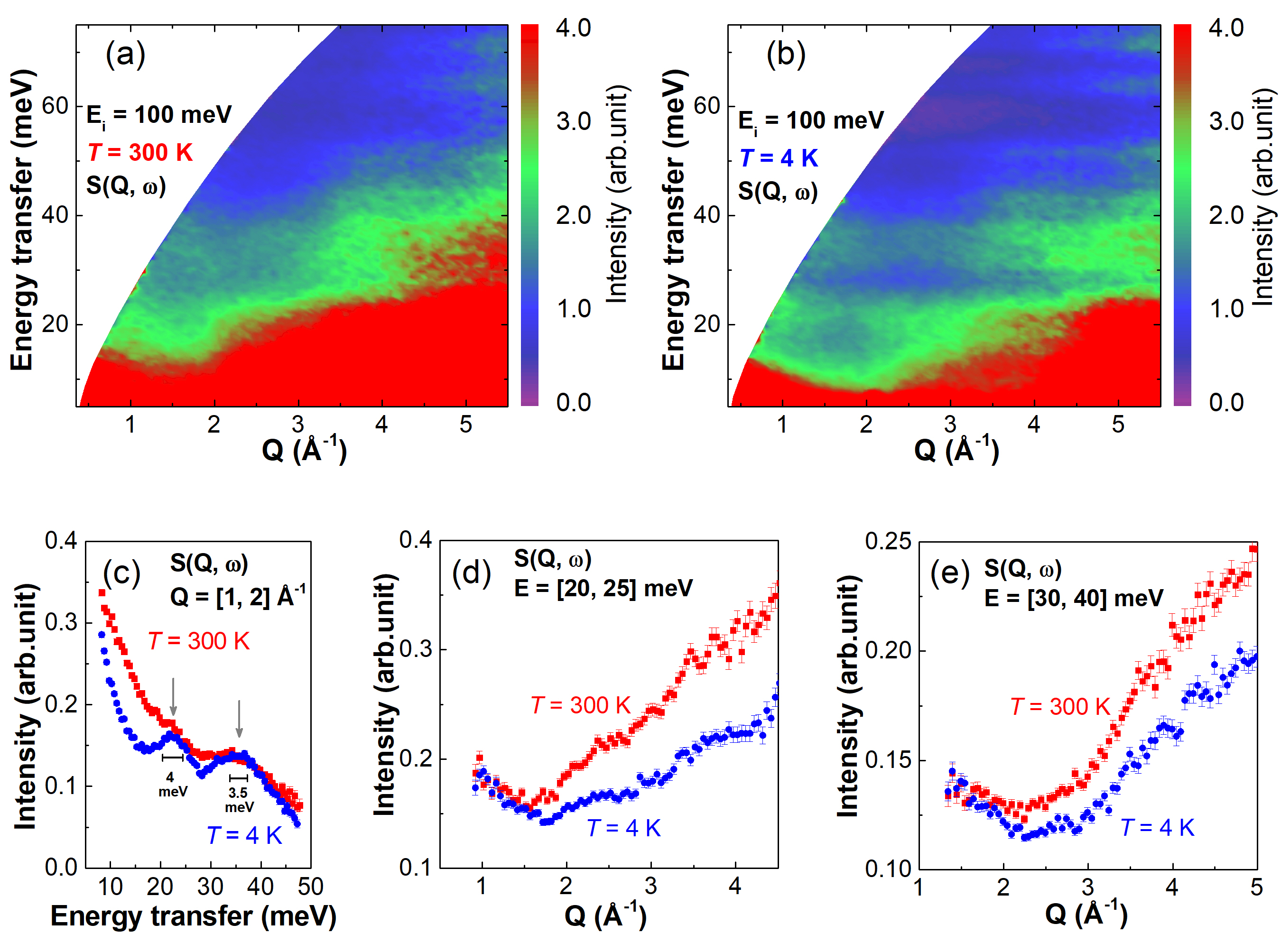}}
\caption{(color online) (a), (b) Color contour plots of the dynamical structure factor $S(Q,\omega)$ multiplied by the magnetic form factor squared $f(Q)^2$ at $T$~$=$~300 K and 4 K, respectively, for the incident energy $E_i$ = 100 meV SEQUOIA data sets. (c) Constant-$Q$ cuts of $f(Q)^2S(Q,\omega)$ at low $Q$ (integration range from 1 to 2~\AA$^{-1}$) with $T$~$=$~4 and 300 K for $E_i$~$=$~100 meV. The gray arrows indicate two candidate magnetic excitations and the horizontal black lines represent instrumental energy resolution at these peak positions. (d), (e) Constant-$E$ cuts of $f(Q)^2S(Q,\omega)$ corresponding to the two peaks in panel (c) show that their intensity increases with decreasing $Q$ and therefore they have a magnetic origin. For simplicity, $f(Q)^2$ is not included in the labels of the figure panels.}
\label{Fig3}
\end{figure*}

To obtain additional evidence for the exotic intermediate spin state of the Ru dimers in \ch{Ba3LaRu2O9}, INS measurements were performed on the SEQUOIA spectrometer with an incident energy $E_i=100$\,meV. Figure \ref{Fig3}(a) and (b) show the color contour plots of the dynamical structure factor $S(Q,\omega)$ multiplied by the magnetic form factor squared $f(Q)^2$ at room temperature (300\,K) and base temperature (4\,K), respectively. The spectra are dominated by the phonon modes in the high momentum transfer ($Q$) regions, which makes it challenging to identify any weak magnetic modes. For this reason, constant-$Q$ cuts of $f(Q)^2S(Q,\omega)$ for both the 4\,K and 300\,K data sets are plotted in Fig.~\ref{Fig3}(c) with a $Q$ integration range of 1 to 2~\AA$^{-1}$. We can now clearly observe two peaks centered at energy transfers $E$~$\approx$~22 and 35\,meV, indicated by gray arrows, corresponding to magnetic excitation candidates. The $Q$-dependence of these two peaks is shown with constant energy cuts plotted in Fig.~\ref{Fig3}(d) and (e) respectively. The intensity of both peaks decreases with increasing $Q$ in the low-$Q$ ($<$~2~\AA$^{-1}$) region and therefore they have a magnetic origin. The persistence of these two modes up to 300~K suggests that they correspond to d-d excitations and not collective magnetic excitations (i.e. spin waves). On the other hand, we identified a third magnetic mode just above the elastic line that is dispersive in nature with the expected temperature-dependence for a spin wave origin. This mode is most clearly observed in lower $E_i$~$=$~25\,meV data and will be discussed in more detail later. 

\begin{table}[!ht]
	\centering
	\caption{Lattice constants and selected structural parameters for Ba$_3$(Y,In,Lu)Ru$_2$O$_9$ \cite{17_ziat} and \ch{Ba3LaRu2O9} extracted from refinements of the NPD data measured at 1.5\,K and 3.5\,K with neutron wavelength $\lambda = 1.54$ \AA.}
	\begin{tabular}{ r l l l l }
		\hline\hline
		\textit{B'}  & In (3.5\,K) & Y (3.5\,K)  & Lu (1.5\,K) & La (1.5\,K) \\[0.5ex]
		\hline
		\textit{a}(\AA)  & 5.7947(1) &	5.8565(1) &	5.8436(1) & 5.95103(14)\\
		\textit{c}(\AA)\  & 14.2738(2) & 14.4589(1) & 14.3978(2) & 15.0087(4)\\
		Ba$_2$-\textit{z} & 0.9116(2) &	0.9075(1)&	0.9084(2)  & 0.8910(2)\\
		Ru-\textit{z}  & 0.1611(1) &	0.1632(1)&	0.1620(1) & 0.16458(16)\\
		O$_1$-\textit{x}  & 0.4874(5) &	0.4879(4) &	0.4887(5) & 0.4866(5)\\
		O$_2$-\textit{x}  & 0.1712(4) &	0.1758(2) &	0.1741(3) & 0.1787(4)\\
		O$_2$-\textit{z}  & 0.4150(1) &	0.4124(1) &	0.4138(1) & 0.40457(8) \\		
		$\textit{R}_\text{wp}(\%)$  &8.82 &	6.27 &	6.18 & 5.83\\		
		Ru-Ru(\AA)  & 2.538(3) & 2.511(2) &	2.533(3) & 2.564(3)\\
		Ru-O$_1$(\AA)  & 2.001(3) &	2.009(2) &	2.019(2) & 2.034(3)\\		
		Ru-O$_2$(\AA) & 1.956(2) &	1.936(1) &	1.947(2) & 1.902(3)\\		
		Ru-O$_1$-Ru($^\circ$)  & 78.8(1) & 77.4(1) & 77.7(1) & 78.16(15)\\[0.5ex]		
		\hline\hline
	\end{tabular}
	\label{table:1}
\end{table}

Our previous magnetic susceptibility and INS work on Ba$_3$(Y,In,Lu)Ru$_2$O$_9$ \cite{17_ziat} revealed a single d-d excitation in all three cases with an energy transfer ranging between 31.5 and 34 meV, which we identified as a molecular transition from the $S_\text{tot}$~$=$~1/2 ground state to the $S_\text{tot}$~$=$~3/2 excited state within a quasi-molecular orbital picture. This interpretation is consistent with the neutron scattering selection rule $\Delta S$~$=$~0,$\pm$1 \cite{10_furrer}. Although these excited modes were not resolution-limited, the broadening may arise from a finite amount of zero field splitting. The current INS data on Ba$_3$LaRu$_2$O$_9$ provides an interesting contrast, as two d-d excitations are observed in a similar energy regime. This observation is consistent with an $S_\text{tot}$~$=$~3/2 electronic ground state for the Ru dimers if these two modes represent transitions to the $S_\text{tot}$~$=$~1/2 and $S_\text{tot}$~$=$~5/2 manifolds, as these excitations are both allowed by selection rules. We assign the lower and upper modes to the $S_\text{tot}$~$=$~1/2 and 5/2 transitions, respectively, as the former is nearly resolution-limited while the latter exhibits increased broadening expected to arise from significant zero field splitting of an $S_\text{tot}$~$=$~5/2 state. We also note that this assignment is consistent with the high-temperature magnetic susceptibility data described above and the previous determination that the nearly isostructural systems Ba$_3$(Y,In,Lu)Ru$_2$O$_9$ host $S_\text{tot}$~$=$1/2 electronic ground states. 

\begin{figure*}[ht]
\centering 
\scalebox{0.75}{\includegraphics{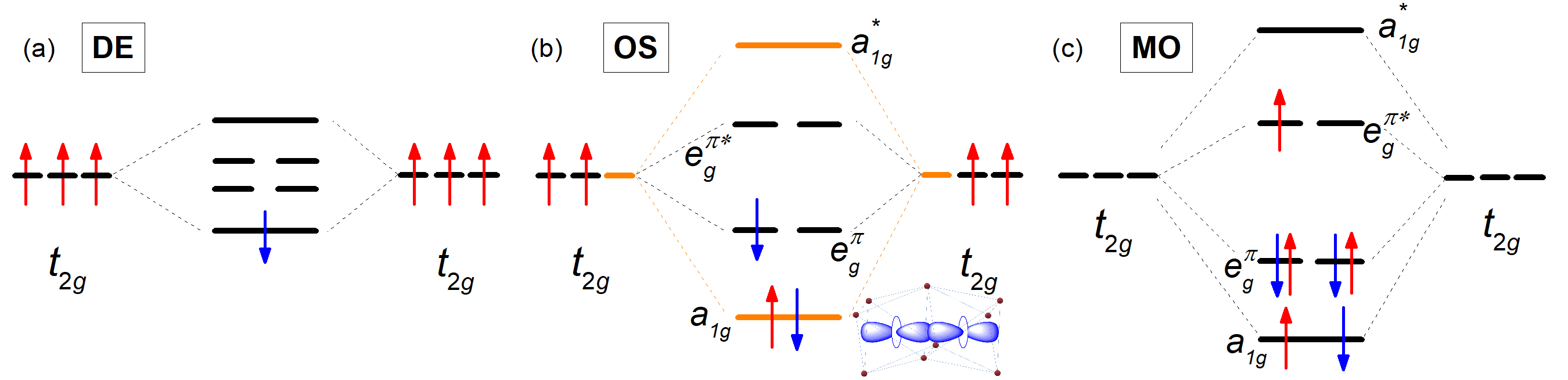}}
\caption{(color online) Schematics of the electronic ground states for Ru dimers with seven electrons based on face-sharing octahedra in the (a) double exchange (DE) limit, (b) orbital-selective (OS) regime, and (c) molecular orbital (MO) limit. The inset of panel (b) shows the $a_{1g}$ orbitals for each Ru dimer. The large direct overlap of the $a_{1g}$ orbitals, combined with the significantly reduced orbital overlap for the $e_g^\pi$ manifold, generates an orbital-selective state in \ch{Ba3LaRu2O9}.}	
\label{Fig4}
\end{figure*}

The orbital diagram for Ba$_3$(Y,In,Lu)Ru$_2$O$_9$, with seven valence electrons per Ru dimer, has been discussed previously and consists of a lower-energy $a_{1g}$ bonding level and a higher-energy $e_g^\pi$ bonding level \cite{17_ziat} that are made up of linear combinations of the atomic $d$-orbitals \cite{15_kugel}. In these cases, the extremely short Ru-Ru distances arising from the face-sharing octahedral geometry of the Ru dimers and the spatially-extended 4$d$ orbitals lead to the low-spin (i.e. $S_\text{tot}$~$=$~1/2) molecular orbital diagram illustrated in Fig.~\ref{Fig4}(c), rather than the high-spin (i.e. $S_\text{tot}$~$=$~5/2) double exchange scenario shown in Fig.~\ref{Fig4}(a) typically expected for molecular magnets based on 3$d$ transition metals. To gain some insight into why the Ru dimers realize a different electronic ground state in \ch{Ba3LaRu2O9}, we revisited the low-temperature crystal structure of this system using the HB-2A powder diffractometer with a neutron wavelength of 1.54~\AA. Our new refinement results for \ch{Ba3LaRu2O9} are presented in Table I and compared to our previous work on Ba$_3$(Y,In,Lu)Ru$_2$O$_9$ \cite{17_ziat}. While we find broad agreement with earlier diffraction work \cite{02_doi, 13_senn}, we note that our low-temperature Ru-Ru and Ru-O$_1$ distances are slightly larger for \ch{Ba3LaRu2O9} compared to Ba$_3$(Y,In,Lu)Ru$_2$O$_9$. Since the $a_{1g}$ orbitals are aligned and overlap directly, as shown in the inset of Fig.~\ref{Fig4}(b), the $a_{1g}$ orbital hopping $t_c$ is most effectively tuned with the Ru-Ru distance. The situation is quite different for the $e_g^\pi$ orbitals, where the reduced direct overlap ensures that the $e_g^\pi$ orbital hopping $t_d$ is smaller and determined by both the Ru-Ru and Ru-O$_1$ distances. It appears that the seemingly subtle differences in the Ru-Ru and Ru-O$_1$ distances in \ch{Ba3LaRu2O9} and Ba$_3$(Y,In,Lu)Ru$_2$O$_9$ are still large enough to effectively tune $t_d$ and generate a spin state transition in this family of materials. This scenario provides a natural explanation for the intermediate spin state of the Ru dimers in \ch{Ba3LaRu2O9}, as it can arise from the orbital diagram presented in Fig.~\ref{Fig4}(b). Due to the large $t_c$~$>$~$J_H$ and the comparatively smaller $t_d$, only the $a_{1g}$ manifold participates in molecular bonding. Presumably, the other five electrons engage in a double exchange process to generate the $S_\text{tot}$~$=$~3/2 spin degree of freedom. 

\subsection{(II) Collective static magnetic properties}

With the electronic ground state of the Ru dimers in \ch{Ba3LaRu2O9} firmly established as $S_\text{tot}$~$=$~3/2 due to orbital-selective behavior, we now examine the collective static magnetic properties of this molecular magnet. There are two previous reports on this topic \cite{02_doi, 13_senn}, but several open questions remain. The initial X-ray diffraction and bulk characterization study reveals two possible magnetic transitions in the specific heat at $T_{c1}$~$\approx$~6 K and $T_{c2}$~$\approx$~22 K, while the magnetic susceptibility shows a clear peak at $T_{c1}$ and a very subtle bump at $T_{c2}$. Follow-up neutron powder diffraction work using the WISH spectrometer at ISIS revealed magnetic Bragg peaks for $T$~$\le$~10 K, but no precise magnetic transition temperature was reported so the origin of $T_{c1}$ and $T_{c2}$ remains unknown. Furthermore, the magnetic Bragg peaks observed in the NPD data were modeled within a local moment picture with intradimer ferromagnetic exchange that was noted to be unusual and the magnetic structure could only be explained by a model consisting of two irreducible representations with a confounding moment direction. 

\begin{figure*}[ht]
\centering
\scalebox{0.75}{\includegraphics{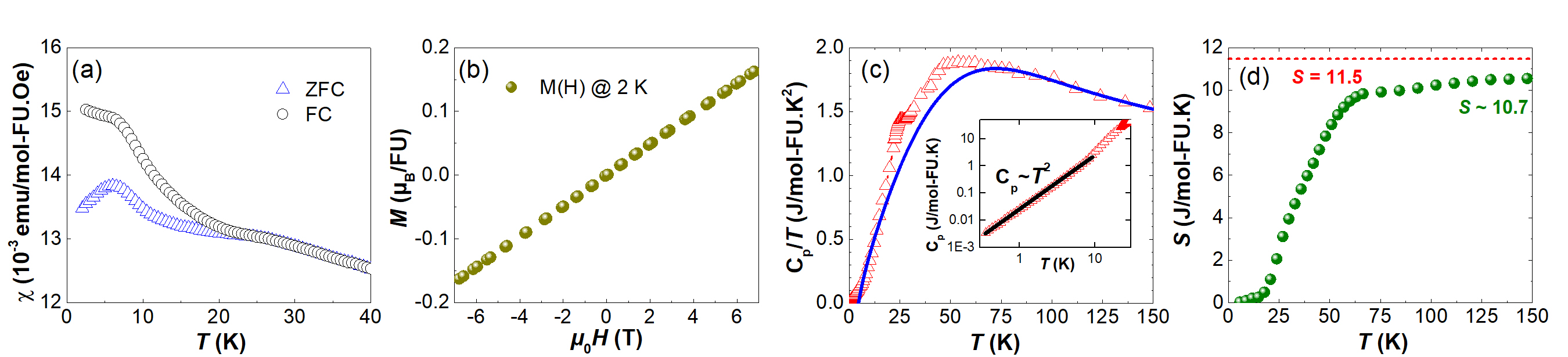}}
\caption{(color online) (a) The low-temperature magnetic susceptibility of \ch{Ba3LaRu2O9} shows a zero-field-cooled / field-cooled divergence below $T_{c2}$~$\approx$~25\,K. (b) The isothermal magnetization measured at 2\,K shows a linear field dependence, which is consistent with an antiferromagnetic ground state. (c) The temperature-dependence of the specific heat $C_p$ shows a clear transition at $T_{c2}$~$\approx$~25\,K. The blue curve is our best fit to the lattice contribution using a Thirring model. We attribute the additional contribution of the magnetic $C_p$ to a Schottky anomaly arising from a small zero field splitting of the $S_\text{tot}$~$=$~3/2 state. The inset depicts the low-temperature $C_p$ on a log-log scale to highlight the $T^2$ dependence in this regime. (d) The magnetic entropy of \ch{Ba3LaRu2O9}, plotted as a function of temperature, approaches the expected value of 11.5 J/mol-K for a $S_\text{tot}$~$=$~3/2 state when warming above the magnetic ordering temperature and the Schottky anomaly. }
\label{Fig6}
\end{figure*}

We performed a series of measurements to explore these issues. First, we present dc magnetic susceptibility data up to 40 K on our \ch{Ba3LaRu2O9} polycrystalline samples in Fig.~\ref{Fig6}(a). We find a zero-field-cooled (ZFC) / field-cooled (FC) divergence that onsets below $T_{c2}$~$\approx$~25~K and a broad peak at $T_{c1}$~$\approx$~6 K in both the ZFC and FC data. These findings are in broad agreement with previous work \cite{02_doi}, although our $T_{c2}$ value is slightly higher. We also plot the magnetization as a function of field at 2 K in Fig.~\ref{Fig6}(b) and we find a linear response which is indicative of a collinear antiferromagnetic ground state. Next, we show heat capacity data over a much wider temperature range than published previously \cite{02_doi} in Fig.~\ref{Fig6}(c). Interestingly, this data shows a clear anomaly at $T_{c2}$~$\approx$~25 K and a Schottky anomaly centered around 50~K, but no obvious feature around $T_{c1}$. We also note that the lowest temperature data measured between 300 mK and 10 K exhibits a $T^2$-dependence as shown in the Fig.~\ref{Fig6}(c) inset. This $T$-dependence can arise from different origins, including a spin wave contribution from a gapless quasi-two-dimensional antiferromagnet, and will be discussed in more detail later. The magnetic entropy was extracted from the $C_p$ data after subtracting the lattice contribution that was approximated by a Thirring model \cite{13_thirring, 89_thirring, 05_thirring} and the result is presented in Fig.~\ref{Fig6}(d). We find that the entropy recovered up to 150~K ($S$~$=$~10.7 J/mol FU-K) is only slightly lower than expectations for a $S_\text{tot}$~$=$~3/2 molecular degree of freedom ($S$~$=$~11.5 J/mol FU-K), which is consistent with the Ru dimer electronic ground state that we described above. This result also suggests that the Schottky anomaly arises from a small zero field splitting of the $S_\text{tot}$~$=$~3/2 state. 
 
\begin{figure}[ht]
\centering
\scalebox{0.54}{\includegraphics{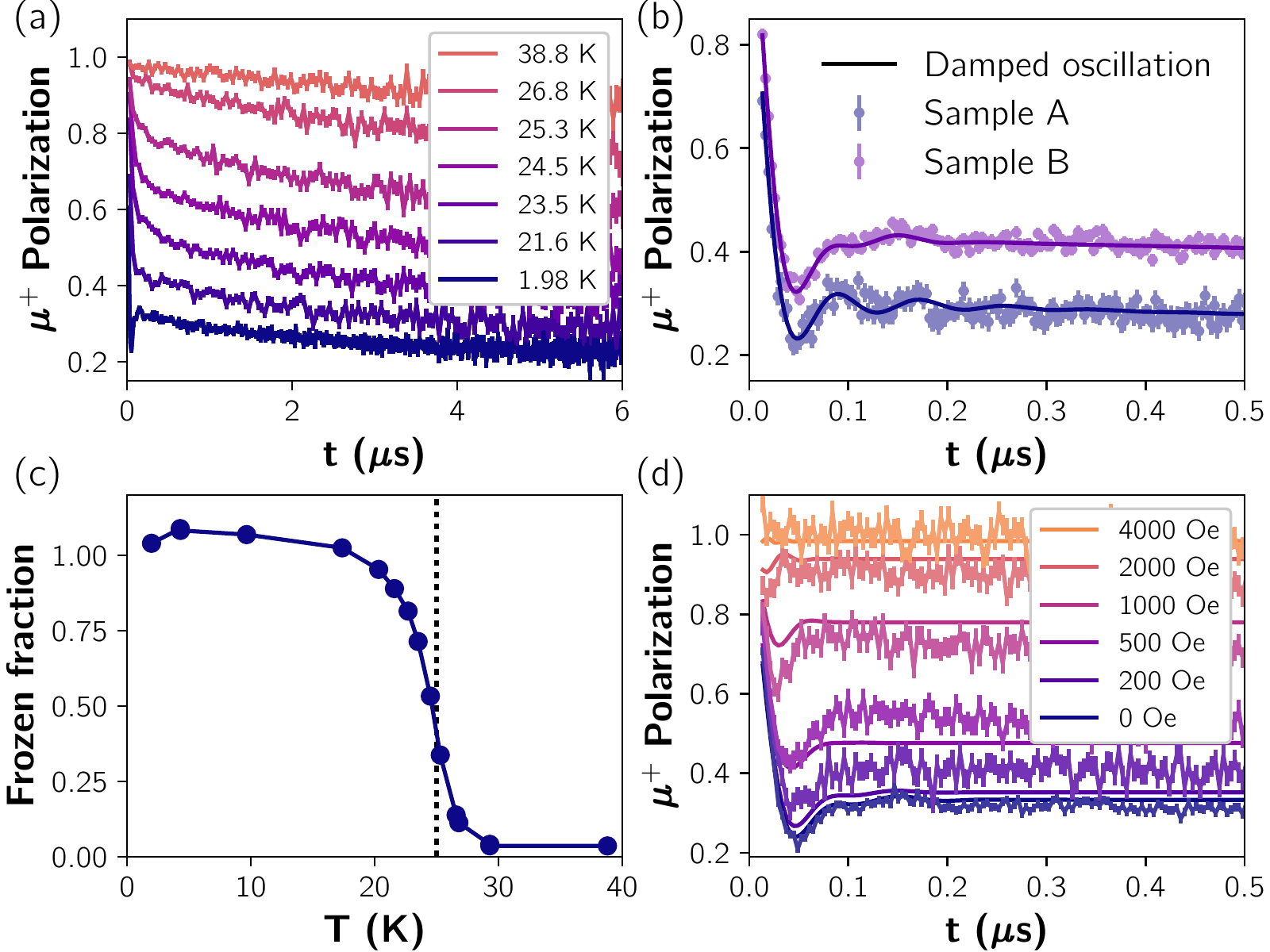}}
\caption{(color online) (a) Temperature-dependence of the muon spin polarization in zero field. (b) Oscillations observed in the zero-field polarization at short times for two different samples, with the sample B data shifted upwards to improve clarity. The data fit well to the sum of two damped cosines plus a 1/3 relaxing tail. To within the uncertainty of the fit parameters, both samples exhibit the same oscillation frequencies. (c) Frozen (ordered) volume fraction vs temperature obtained from the zero field measurements. The magnetic transition temperature is indicated by a vertical dashed line. (d) Longitudinal field scan of the muon spin polarization at $T$~$=$~1.5~K, along with simulation results described in the text shown as solid curves.}
\label{musrFigure}
\end{figure}

Since we were not able to identify a definitive origin for the $T_{c1}$ magnetic transition in \ch{Ba3LaRu2O9} from our bulk characterization measurements, we performed muon spin relaxation ($\mu$SR) on our samples. This technique is extremely sensitive to local magnetic fields and magnetic volume fractions, so it can be used to readily differentiate between scenarios where the $T_{c1}$ transition arises from a small magnetic impurity phase or a magnetic structure change intrinsic to \ch{Ba3LaRu2O9}. Muon spin polarization (i.e. $\mu^+$ polarization) plotted as a function of time at various temperatures is shown in Fig.~\ref{musrFigure}(a) and reveals clear evidence of static magnetism appearing below a temperature of roughly 25\,K, which is in excellent agreement with our bulk characterization data. Closer inspection of the lowest temperature data reveals highly-damped oscillations, as shown in Fig.~\ref{musrFigure}(b), indicating long-range order with appreciable decoherence (short $1/T_2$). The polarization could be successfully fit with the following function:
\begin{equation} P(t) = \frac{2}{3}\sum_{i=1}^2 a_i e^{-t/T_{2i}}\cos(\omega_i t) + \frac{1}{3} e^{-t/T_1},
\end{equation}
where $1/T_2$ is the dephasing or decoherence rate and $1/T_1$ is the spin-lattice relaxation rate. Two different frequencies were used, which likely correspond to muon stopping sites near the two inequivalent oxygen sites in the structure since $\sim$~1\AA~muon-oxygen bonds are commonly found in oxides \cite{83_holzschuh}. It was not possible to successfully fit the data with the Koptev-Tarasov function~\cite{11_yaouanc}, which accounts for damping of oscillations through significant inhomogeneity of the internal fields, rather than decoherence. Similar results were obtained for two samples studied (A and B). Slight discrepancies are observed, but the oscillation frequencies are the same within the uncertainty on the fitting parameters. More precisely, in sample A we obtained $\omega_{1A} = 72\pm3$ $\mu$s$^{-1}$ and $\omega_{2A} = 40\pm 13$ $\mu$s$^{-1}$. In sample B, $\omega_{1B} = 79\pm8$ $\mu$s$^{-1}$ and $\omega_{2B} = 42\pm 5$ $\mu$s$^{-1}$. Since this data for the two samples is in excellent agreement, we focus on the sample A measurements in the rest of this section. 

To identify the onset of the magnetic order and hence the transition temperature, we have fit the slowly-relaxing part of the zero-field polarization at $t>0.2$ $\mu$s for sample A to a simple exponential function, $P_\mathrm{ZF} = (1 - 2f/3)e^{-t/T_1}$, where $f$ represents the frozen fraction. Here the amplitude of the slowly-relaxing component contains both the non-frozen fraction of the sample and the 1/3 tail from the static fraction. The temperature-dependence of the ordered volume fraction extracted from this analysis is shown in Fig.~\ref{musrFigure}(c) and essentially consistent with a fully-ordered sample.

In Fig.~\ref{musrFigure}(d), a longitudinal field scan of the muon spin polarization is shown and demonstrates that $B_L = 4000$\,Oe is sufficient to completely decouple the muon spins from internal magnetic fields (which are at most around 850\,Oe).  A simulation of the expected longitudinal field behavior for each muon stopping site was performed using the following equation:
\begin{equation} 
\begin{split} P(t) = & \frac{1}{2}\int_0^\pi d\xi \sin\xi \sin^2\left[ \theta(\xi)\right] \cos\left[ B(\xi)\gamma t\right]e^{-t/T_{2i}}  \\
+ & \frac{1}{2}\int_0^\pi d\xi \sin\xi \cos^2\left[ \theta(\xi) \right]e^{-t/T_1},
\end{split}
\end{equation}
with the magnitude of the magnetic field given by
\begin{equation} B = \sqrt{(B_L + B_i\cos\xi)^2 + B_i^2\sin^2\xi}, \end{equation}
and with $\cos\theta = (B_L + B_i\cos\xi)/B$ and $\sin\theta = (B_i/B)\sin\xi$. Here, $B_i$ represents the internal field for the $i$th muon stopping site in the absence of applied field. While this simulation [see Fig.~\ref{musrFigure}(d)] is not entirely successful, it is also quite an oversimplification. Here we only consider the effects of the addition of the longitudinal field onto a unique (but randomly-oriented) internal field with linewidth and relaxation effects added afterwards in an \emph{ad hoc} fashion. The main deviations from theory are at low fields where the linewidth is comparable to the applied field and our model is particularly crude. The fields at which complete decoupling is achieved are well-captured by the model and overall this analysis is fully consistent with the notion of static magnetism in \ch{Ba3LaRu2O9}. 

These $\mu$SR measurements provide support for two important conclusions. Firstly, the highest frequency observed in these samples (72$\pm3$ $\mu$s$^{-1}$ for sample A and 79$\pm8$ $\mu$s$^{-1}$ for sample B) is roughly six times higher than the highest frequency observed in the isostructural material \ch{Ba3LuRu2O9}~\cite{17_ziat}. This finding can be easily rationalized with the higher local magnetic fields expected for the $S_\text{tot}$~$=$~3/2 Ru dimer ground state discussed above. Naively, one would expect a factor of three increase in internal field, but this expectation completely neglects the rather different orbital configurations that will result in these two distinct situations. Secondly, there is no abrupt change in the magnetic volume fraction of the data at $T_{c1}$. This suggests that the 6 K magnetic transition arises from a small magnetic impurity in the sample, and in fact the 12-L perovskite \ch{Ba4LaRu3O12} is known to order at this temperature \cite{10_doi}. We note that the true ordering temperature of \ch{Ba3LuRu2O9} is significantly higher than its isostructural counterparts Ba$_3$(Y,In,Lu)Ru$_2$O$_9$ \cite{02_doi, 17_ziat} and this is fully consistent with expectations for an electronic ground state with a larger spin. 

\begin{figure*}[ht]
\centering 
\scalebox{0.75}{\includegraphics{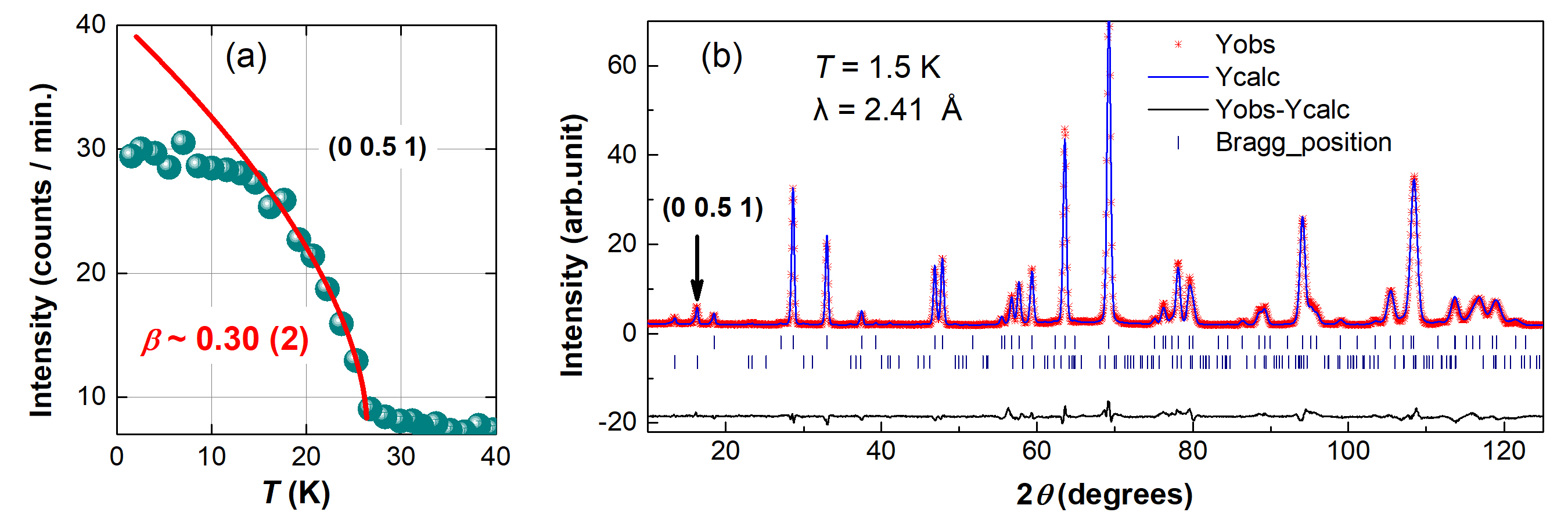}}
\caption{(color online) (a) The temperature-dependence of the (0 0.5 1) magnetic peak intensity. A power law fit performed just below the transition temperature reveals a critical exponent of 0.30(2), which is consistent with a three-dimensional universality class. (b) The refined neutron powder diffraction pattern at 1.5 K. The data is plotted with red symbols, the combined structural and magnetic model is superimposed with a blue curve, the ticks below the pattern show expected Bragg peaks for the structural and magnetic phase, and the difference pattern is indicated by a solid black curve. The index of the strongest magnetic peak is (0 0.5 1).  }
\label{Fig7}
\end{figure*}

In order to obtain confirmation of the true magnetic transition temperature and to better understand the critical behavior at this transition, complementary elastic neutron scattering measurements were performed on a polycrystalline sample of \ch{Ba3LaRu2O9} using the HB-1A triple axis spectrometer at HFIR. Fig.~\ref{Fig7}(a) shows the temperature dependence of the intensity for the strongest magnetic peak, which can be indexed by (0 0.5 1) as explained below. A simple power law was applied to fit the peak intensity near the transition temperature:
\begin{equation}
I = I_0\left(1-\frac{T}{T_N}\right)^{2\beta},
\end{equation}
where $T_N$~$=$~$T_{c2}$ is the Ne\'{e}l temperature and $\beta$ is the critical exponent of the order parameter (OP). The fitting result yields $T_N$~$=$~26(1)\,K and no sharp change in the intensity is detected around 6 K, which is consistent with our other measurements described above. We also find that $\beta$~$=$~0.30(2), which is close to the values expected for a three-dimensional universality class ($\beta_\text{3D, Ising}$~$=$~0.326 and $\beta_\text{3D, Heisenberg}$~$=$0.345) and much larger than expected for a quasi-2D Ising model ($\beta_\text{2D, Ising}$~$=$~0.125). This result is surprising since a $T^2$-dependence for the low-temperature specific heat often arises from spin wave contributions of a gapless, quasi-2D antiferromagnet. 

With the true magnetic transition temperature of \ch{Ba3LuRu2O9} now established, we return to the open questions surrounding the magnetic structure measured previously \cite{13_senn}. We collected neutron powder diffraction data on polycrystalline \ch{Ba3LaRu2O9} at 1.5 and 40 K using the HB-2A powder diffractometer with a neutron wavelength of 2.41~\AA. The 1.5 K diffraction pattern is similar to the previous measurements of Senn {\it et al} \cite{13_senn} and consists of both nuclear and magnetic Bragg peaks, as shown in Fig.~\ref{Fig7}(b). The 1.5 K data can be refined in the $P6_3/mmc$ space group, as noted above, and the magnetic peaks observed can be indexed with the same propagation vector $\vec{k}$~$=$~(0 0.5 0) identified previously. To model the magnetic structure we first performed a symmetry analysis using SARAh \cite{00_wills}. Assuming a second order phase transition at $T_{N}$, the most likely magnetic models should correspond to one of the eight irreducible representations described in Ref.~\cite{13_senn}. However, we found that none of these models could fully explain our data, and therefore we also tried linear combinations of them. Ultimately, we find that the best magnetic refinement of the 1.5~K diffraction pattern is achieved by using the same $\Gamma_3 + \Gamma_5$ model as before \cite{13_senn}. The $\Gamma_3$ component is required to explain the magnetic intensity at the (-1 0.5 0) position, which is not captured by the $\Gamma_5$ model. To estimate the ordered moment size, we used a local moment model with the Ru$^{5+}$ magnetic form factor that has been reported elsewhere \cite{03_parkinson}. The refined moment sizes per Ru ion at $T$~$=$~1.5~K are $m_b=1.23(9)\,\mu_B$, $m_c=0.5(1)\,\mu_B$, and $m_\text{tot}=1.3(1)\,\mu_B$. These values are consistent with the reported ordered moment sizes of $m_b=1.3(1)\,\mu_B$, $m_c=0.6(2)\,\mu_B$, and $m_\text{tot}=1.4(2)\,\mu_B$ reported previously \cite{13_senn} and close to the expectation of $m_\text{tot}=3\,\mu_B$ for a $S_\text{tot}$~$=$~3/2 Ru dimer electronic ground state. We also note that the orbital-selective Mott state for the Ru dimers naturally describes the ferromagnetic intradimer coupling revealed from the analysis of the NPD data, but the refined moment direction is difficult to understand. A schematic of the refined magnetic structure is presented in Figs.~\ref{Fig1}(a) and (c); this is the collinear stripe spin configuration that is predicted for a triangular lattice with a significant in-plane next nearest neighbor exchange interaction $J_\text{NNN}$ (i.e. $J_\text{NNN}/J_\text{NN}>0.125$ \cite{11_seabra}). 

\subsection{(III) Collective spin dynamics}

\begin{figure}[ht]
\centering 
\scalebox{0.5}{\includegraphics [width = 0.8\textwidth] {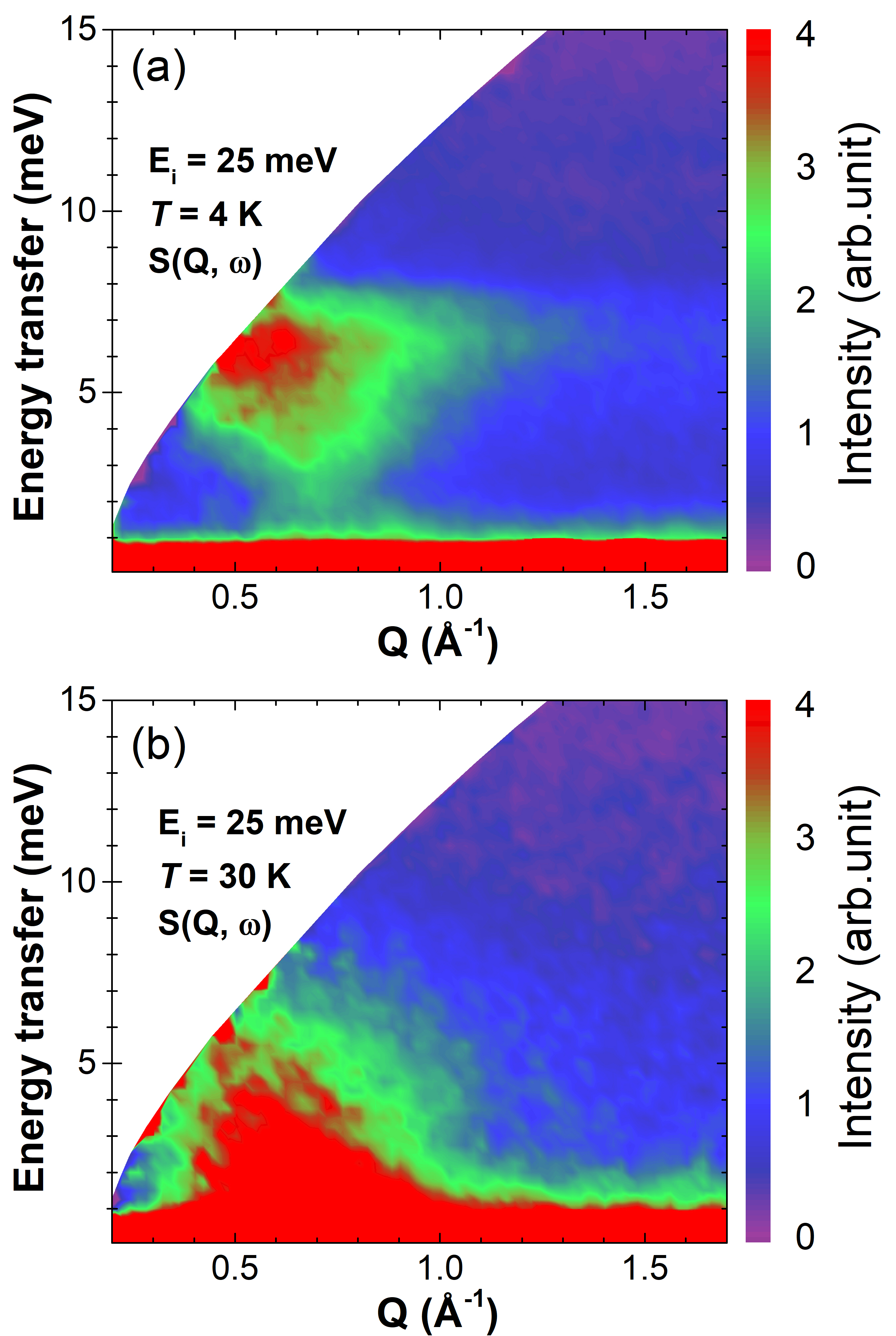}}
\caption{(color online) (a), (b) Color contour plots of the dynamical structure factor $S(Q,\omega)$ multiplied by the magnetic form factor squared $f(Q)^2$ at $T$~$=$~4 K and 30 K, respectively, for the incident energy $E_i$ = 25 meV SEQUOIA data sets. The spectral weight of the inelastic scattering observed at 30~K shifts up to higher energy transfers below $T_{N}$, which is consistent with a gapped spin wave mode. For simplicity, $f(Q)^2$ is not included in the labels of the figure panels. }
\label{Fig8}
\end{figure}

We now return to the spin wave mode measured with inelastic neutron scattering using the SEQUOIA spectrometer. This excitation was best observed by collecting data with an incident energy $E_i$~$=$~25~meV, which is presented in Fig.~\ref{Fig8}(a) and (b) at $T$~$=$~4 and 30 K respectively. Notably, there is a strong band of inelastic scattering at 30 K centered above the (0 0.5 0) and (0 0.5 1) magnetic Bragg peaks that shows a significant shift in spectral weight up to higher energy transfers at 4 K. This phenomenon has been observed in many other ordered systems when cooling below $T_N$ \cite{13_carlo, 14_aczel, 16_taylor} and may suggest that the spin wave spectrum of Ba$_3$LaRu$_2$O$_9$ is gapped. However, the spin wave contribution to the low-temperature specific heat would then show an activated behavior rather than a $T^2$ dependence. This discrepancy, combined with the inconsistent universality class conclusions obtained from the specific heat and the HB-1A neutron scattering data, suggest that the $T^2$-dependence of the low-temperature specific heat does not arise from a spin wave contribution. 

We also considered modeling the spin wave data with a magnetic Hamiltonian, but we failed to find a simple model that could explain the moment direction obtained from neutron powder diffraction. For this hexagonal crystal structure, Heisenberg models with zero field splitting (i.e. single ion anisotropy) $D$ can only produce moments in the ab-plane ($D$~$>$~0) or along the c-axis ($D$~$<$~0), but not somewhere in between these two extremes. To obtain the correct ground state found experimentally, it appears that the addition of exchange anisotropy is essential, but this consideration goes beyond the scope of our work on a powder sample. Single crystal inelastic neutron scattering measurements on \ch{Ba3LaRu2O9} will be invaluable for elucidating the magnetic Hamiltonian of this system. 

\subsection{(IV) Structural and spin state transitions under pressure}

\begin{figure*}[ht]
\centering 
\scalebox{0.75}{\includegraphics{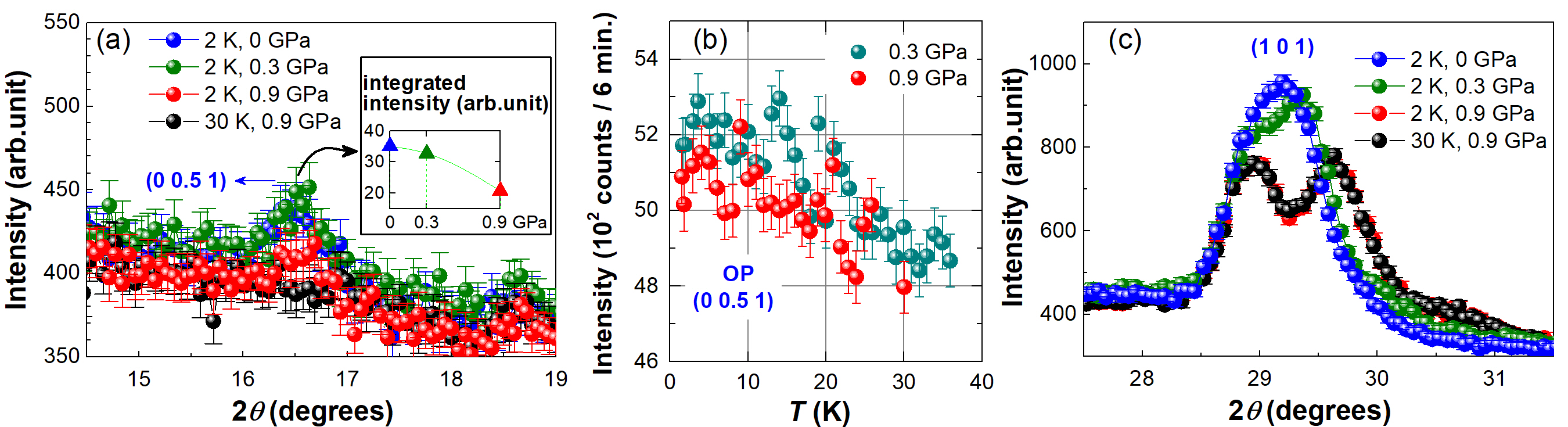}}
\caption{(color online) (a) A close-up view of the (0 0.5 1) magnetic Bragg peak as a function of different temperatures and applied pressures. At a temperature of 2 K, this peak is significantly suppressed at an applied pressure P~$=$~0.9~GPa. Due to a small background shift in the data with increasing pressure, this effect can be best observed by tracking the integrated intensity of this peak as a function of pressure (see the inset). (b) The (0 0.5 1) peak intensity as a function of temperature for P~$=$~0.3 and 0.9 GPa. Although the magnetic transition temperature shows very little change with increasing pressure, the peak intensity is noticeably suppressed by 0.9 GPa, which is indicative of a spin state transition in this pressure regime. (c) A close-up view of the hexagonal (1 0 1) Bragg peak, which splits into three peaks even at P~$=$~0.3~GPa.  }
\label{Fig10}
\end{figure*}

One way to gain additional insight into mechanisms leading to the $S_\text{tot}$~$=$~3/2 molecular ground state in \ch{Ba3LaRu2O9} is to apply external stimuli \cite{12_brinzari,14_oneal}. Since orbital hybridization should increase with decreasing intradimer Ru-Ru distance, our hypothesis is that a small amount of pressure may induce a spin state transition from $S_\text{tot}$ = 3/2 to $S_\text{tot}$ = 1/2 \cite{14_oneal}. Such a transition is expected to take place with a reduction in the Ru ordered moment from 3\,$\mu_B$ to 1\,$\mu_B$ per cluster. We performed a neutron diffraction experiment at HB-2A in a 1 GPa Cu-Be pressure cell to search for evidence of such a high $\rightarrow$ low spin state transition by tracking the intensity of the strongest magnetic Bragg peak under compression. We find significant suppression in the intensity of this peak when the pressure is increased from 0.3 to 0.9~GPa, as shown in Fig.~\ref{Fig10}(a) and (b), but no significant change in the magnetic transition temperature. Due to a background shift with increasing pressure, we plot the pressure-dependence of the integrated intensity for this peak in the Fig.~\ref{Fig10}(a) inset. We also searched for new magnetic Bragg peaks at 0.9\,GPa by collecting diffraction patterns over a wide angular range at both 2 and 30~K, but none were found. Suppression of this magnetic peak under compression is therefore consistent with a low-lying spin state transition rather than a magnetic structure change in this material. Notably, the crystal structure also appears to be modified at a lower pressure of 0.3~GPa, as the hexagonal (101) peak near $2\theta$~$=$~29\textdegree~splits into three peaks, as shown in Fig.~\ref{Fig10}(c). The highest crystal symmetry consistent with this three-fold peak splitting is monoclinic. Unfortunately, our data quality is insufficient for refining the crystal structure of this material under compression due to the high, structured background of the pressure cell and the significant neutron beam attenuation through the cell. 

\begin{figure*}[ht]
\centering 
\scalebox{0.65}{\includegraphics{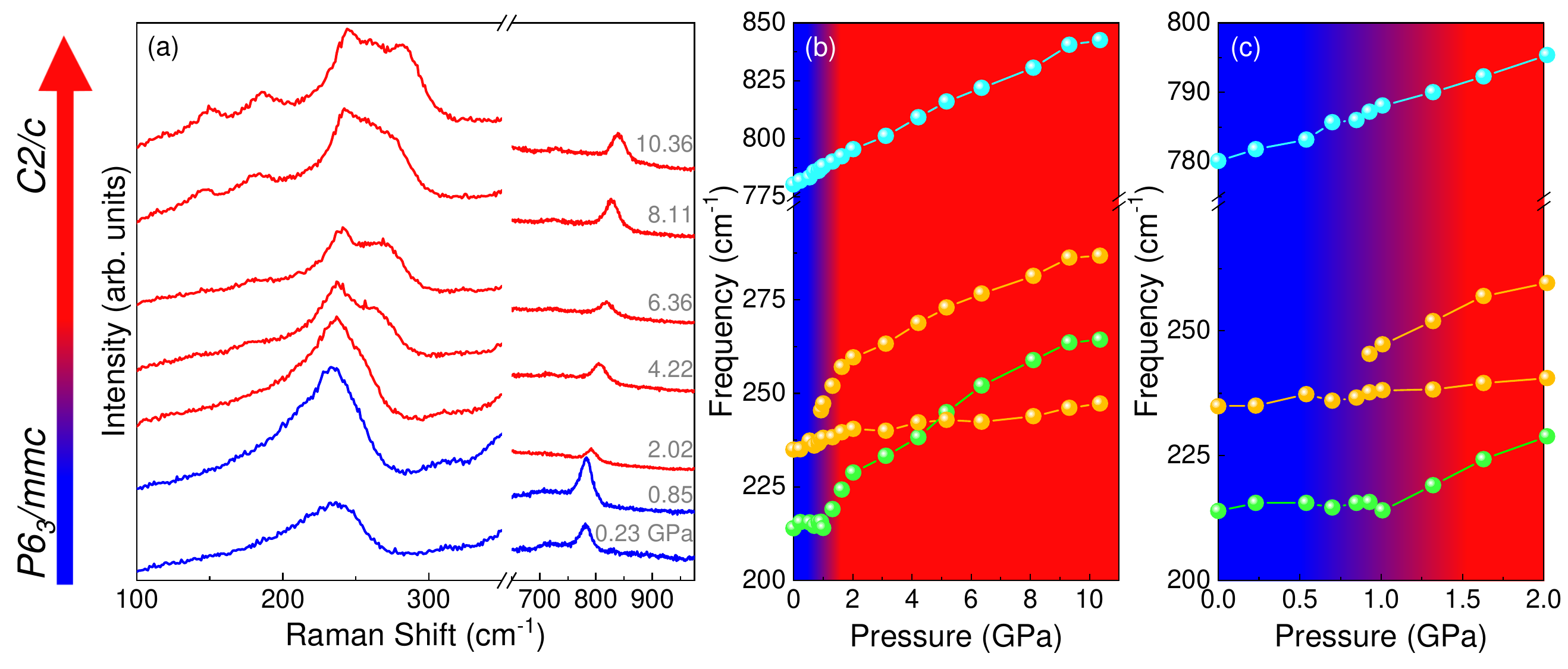}}
\caption{(color online) (a) Raman scattering spectra of \ch{Ba3LaRu2O9} under compression at room temperature. (b) Frequency vs pressure of several characteristic vibrational modes of \ch{Ba3LaRu2O9} up to 10.36 GPa. (c) Close-up view of frequency vs pressure trends in the vicinity of the 0.9 GPa transition. The color scheme denotes the change in crystal structure. }
\label{Raman}
\end{figure*}

In order to gain additional insight into the pressure-driven transitions, we turn to Raman scattering. This method dovetails well with diamond anvil cell techniques and supports tuning the sample over a much wider pressure range. Figure~\ref{Raman}(a) displays the Raman response of \ch{Ba3LaRu2O9} at room temperature under compression. Plots of frequency vs pressure, as shown in Fig.~\ref{Raman}(b) and (c), allow us to track the behavior of individual phonons and identify clear changes at P~$=$~0.9~GPa. While some modes such as the Ru--O stretch at 776 cm$^{-1}$ are insensitive to the 0.9 GPa transition, others (for instance at 215 and 235 cm$^{-1}$) sport inflection points with subsequent hardening as well as strong doublet splitting. This suggests that while the Ru dimer is structurally rigid across the critical pressure, the charge storage layer containing Ba and La is not. This is consistent with findings for structural rigidity of the Ru dimer across the high $\rightarrow$ low spin transition in the bimetallic quantum magnet [Ru$_2$(O$_2$CMe)$_4$]$_3$[Cr(CN)$_6$] \cite{07_shum, 12_brinzari, 14_oneal}. At the same time, the lower frequency modes involving Ba and La motion provide evidence for symmetry breaking across the 0.9 GPa transition. 

A correlation group analysis identifies several candidate subgroups of the $P6_3/mmc$ space group. Recalling that neutron diffraction for $T$~$\le$~30~K constrains the high pressure phase to a monoclinic structure, these subgroups include $C2/c$, $P2_1/m$, and $C2/m$. We begin by considering whether the system will have a primitive or centered lattice type in the high-pressure phase. Our system sports a primitive lattice ($P6_3/mmc$) at ambient conditions and goes through a centered lattice space group ($Cmcm$) on the way to one of the three candidate monoclinic subgroups. We expect that the final symmetry reduction will retain a centered lattice because the face-centered cell is more dense than the primitive lattice and therefore more stable under pressure. The screw operation is also unlikely to remain intact under these conditions. This eliminates $P2_1/m$ and leaves $C2/c$ and $C2/m$ as the remaining candidates. Next we consider whether the high pressure phase contains a reflection or glide plane. Here, we proceed by realizing that reflection is a higher symmetry operation and that pressure tends to break mirror planes. This leaves $C2/c$ as the most probable space group above the critical pressure, which is consistent with previous work identifying a $C2/c$ crystal structure in symmetry-lowering transitions of other 6H-perovskites \cite{12_kimber, 13_senn}. Further work is required to determine whether the 0.9~GPa structural transition at room temperature is coincident with the spin state transition identified at lower temperatures by neutron powder diffraction.


\section{Conclusions}
In conclusion, we have used a combination of bulk characterization, muon spin relaxation, neutron diffraction, and inelastic neutron scattering to identify an intermediate $S_\text{tot}$~$=$~3/2 Ru dimer ground state in \ch{Ba3LaRu2O9} that is generated by orbital-selective Mott insulating behavior at the molecular level. We also find collinear stripe magnetic order below $T_N$~$=$~26(1)~K for these spin-3/2 degrees-of-freedom, which is consistent with expectations for an ideal triangular lattice with significant next nearest neighbor in-plane exchange. Finally, we present neutron diffraction and Raman scattering data under applied pressure that reveal low-lying structural and spin state transitions at modest applied pressures P~$\le$~1\,GPa, which highlights the delicate balance between competing energy scales in this material. Interesting future directions for \ch{Ba3LaRu2O9} include identifying the origin of the $T^2$-dependence of the low-temperature specific heat, determining the magnetic Hamiltonian giving rise to the moment direction of the stripe spin order, solving the high-pressure crystal structure, and carefully mapping out the temperature-pressure phase diagram. Our work highlights the need to develop a comprehensive understanding of the electronic ground state of a heavy transition metal molecular magnet, where large orbital hopping may lead to the breakdown of a simple local moment / double exchange picture, before the collective magnetic properties of the system can be properly identified and characterized.  

\begin{acknowledgments}
Research at the University of Tennessee is supported by the National Science Foundation, Division of Materials Research under award NSF-DMR 1350002 (HDZ) and the Department of Energy, Office of Basic Energy Sciences, Materials Science Division under award DE-FG02-01ER45885 (JLM). JGC is supported by the MOST, NSFC and CAS through projects with Grant Nos. 2018YFA0305700, 11874400, 11921004, and QYZDB-SSW-SLH013. GHW and JM are supported by the MOST and NSFC through projects with Grant Nos. 2016YFA0300501, 11774223 and U1732154. JQ acknowledges research funding obtained from NSERC and the FRQNT. A portion of this research used resources at the High Flux Isotope Reactor and the Spallation Neutron Source, which are DOE Office of Science User Facilities operated by Oak Ridge National Laboratory. 
\end{acknowledgments}

\bibliography{Ba3LaRu2O9_draft_v3}

\end{document}